# Antiferromagnetic iron based magnetoelectric compounds

S. W. Lovesey[1, 2, 3]

[1] ISIS Facility, STFC, Didcot, Oxfordshire OX11 0QX, UK

[2] Diamond Light Source, Harwell Science and Innovation Campus, Didcot, Oxfordshire OX11 0DE, UK

[3] Department of Physics, Oxford University, Oxford OX1 3PU, UK

**Abstract** The Landau free-energy of a compound that benefits from a linear coupling of an electric field and a magnetic field includes a product of the two fields, one polar and time-even and one axial and time-odd. Evidently, the coefficient of the product of fields is unchanged by a simultaneous change in the directions of space and time, a symmetry operation labelled anti-inversion. Invariance with respect to anti-inversion is the defining symmetry of the linear magnetoelectric (ME) effect included in 58 of 122 magnetic crystal classes, 19 of which prohibit higher-order (non-linear) contributions to the free-energy. In ME compounds, expectation values of some atomic magnetic tensors are invariant with respect to anti-inversion. An invariance shared by the Dirac monopole (an element of charge allowed in Maxwell's equations that has not been observed) and a Zel'dovich anapole, also known as a Dirac dipole. From the science of materials perspective, it has been established that Dirac multipoles contribute to the diffraction of x-rays and neutrons. We identify Dirac monopoles in bulk magnetic properties of iron tellurate ($Fe_2TeO_6$) and a spin ladder ($SrFe_2S_2O$). They are visible in the diffraction of light using an iron electric dipole - magnetic dipole absorption event. Both cited compounds present a simple antiferromagnetic configuration of axial dipoles, and their different magnetic crystal classes allow a linear ME effect. However, the Kerr effect is symmetry allowed in the spin ladder and forbidden in iron tellurate. Anapoles are forbidden in iron tellurate and allowed in the spin ladder compound, a difference evident in diffraction patterns fully informed by symmetry. More generally, we identify a raft of Dirac multipoles, and axial multipoles beyond dipoles, visible in future experiments using standard techniques with beams of neutrons or x-rays tuned in energy to an iron atomic resonance. ME invariance imposes a phase relationship between nuclear (charge) and magnetic contributions to neutron (x-ray) diffraction amplitudes. In consequence, intensities of Bragg spots in an x-ray pattern do not change when helicity in the primary beam is reversed. A like effect occurs in the magnetic diffraction of polarized neutrons.

## I. INTRODUCTION

The induction of magnetization by an electric field, and the inverse magnetoelectric (ME) effect, namely, induction of electric polarization by a magnetic field have been studied for a long time. In 1959, Dzyaloshinskii predicted that a linear coupling of the two fields could occur in antiferromagnetic chromium sesquioxide ($Cr_2O_3$) [1]. A year later, the coupling coefficient was experimentally observed to be non-zero in an unorientated crystal below the Néel temperature [2]. Further early work on the ME effect in $Cr_2O_3$ and other antiferromagnetic crystals, such as $Gd_2CuO_4$, $Sm_2CuO_4$, $KNiPO_4$, $LiCoPO_4$ and $BiFeO_3$, is gathered in Refs. [3, 4]. The ME effect is allowed when the magnetic crystal class contains the product of spatial and time inversions, or anti-inversion ($\bar{1}'$), i.e., the two discrete symmetries of space and time do no occur independently. Specifically, the linear ME coupling coefficient in the Landau free-energy is unchanged by a simultaneous change in sign of space and time coordinates. Of 58 classes that allow the linear ME effect, there are 19 that do not allow allow non-linear effects. Examples of the 19 linear types include trigonal $Cr_2O_3$ (magnetic crystal class $\bar{3}'m'$) and tetragonal gadolinium tetraboride $GdB_4$ (4/m'm'm') [5, 6].

Anti-inversion in the crystal class prohibits axial magnetism, which is parity-even (axial) and time-odd (magnetic). Magnetic monopoles are allowed, however. While there is a strong symmetry between electric field and magnetic field in Maxwell's equations, a magnetic charge analogous to electric charge is peculiarly absent. If the equations are symmetrized by the introduction of magnetic charge, this charge must be such that it is reversed by each of the discrete symmetries of parity and time [7]. At an atomic level of detail, a monopole $\langle \mathbf{S} \cdot \mathbf{R} \rangle$ represents magnetic charge, where $\mathbf{S}$ and $\mathbf{R}$ are spin and orbital electronic degrees of freedom, respectively, and angular brackets denote a time-average (expectation value). Such a charge contributes to the diffraction of x-rays utilizing an electric dipole - magnetic dipole event [8, 9, 10]. Notably, a magnetic charge and Dirac's monopole have identical discrete symmetries [7, 11]. An anapole is the next member along in a family of electronic Dirac multipoles and it is equivalent to a dipole $\langle (\mathbf{S} \times \mathbf{R}) \rangle$ and a like orbital entity [12, 13, 14]. The Dirac monopole and the anapole represent the scalar (rank K = 0) and dipole (K = 1) in a decomposition of the third-rank ME tensor coupling coefficient. Anapoles contributes to the Bragg diffraction of x-rays and neutrons, along with higher-order multipoles, e.g., a quadrupole (K = 2) in the decomposition of the ME coefficient [15, 16, 17]. Compton scattering of x-rays is another technique with potential to observe them [18, 19]. Advanced simulations of electronic structure are established methods by which to derive estimates of Dirac multipoles [20, 21, 22].

Looking to the future, we calculate Bragg diffraction patterns for two antiferromagnetic iron based magnetoelectric materials specified in Table I. The configuration of axial magnetic dipoles in iron tellurate ($Fe_2TeO_6$), depicted in Fig. 1, was established in 1968 [23], and its magnetoelectric properties are firmly established [24-27]. Not so for the spin ladder compound $SrFe_2S_2O$, to the best of our knowledge [28]. Axial magnetic dipoles for the spin ladder are depicted in Fig. 2, and the same configuration has been established for $SrFe_2Se_2O$ [28]. All mentioned magnetic configurations have a propagation vector = (0, 0, 0), and the magnetic space groups are centrosymmetric.

Anti-inversion among the elements of symmetry in a magnetic crystal class is a profound influence on diffraction amplitudes. For Bragg diffraction of x-rays enhanced by an atomic resonance, charge and magnetic contributions to scattering amplitudes are in phase [30]. Thus, there is no interference between charge and magnetic contributions to Bragg diffraction patterns gathered from a ME compound. Moreover, coupling to helicity in primary x-rays is forbidden, with no difference in the intensity of a Bragg spot observed with opposite handed x-rays. In contrast, anti-inversion invariance imposes a 90º phase shift between nuclear and magnetic contributions in neutron scattering, and they are in quadrature in the intensity of a Bragg spot [5, 31, 32]. A corollary is that the classical polarized neutron diffraction technique, using a departure from unity of the ratio of intensities for primary neutron beams of opposite polarization, is not available for ME compounds [31, 33, 34, 35]. In their measurement of the magnetization distribution in $Cr_2O_3$, Brown *et al*. exploited spherical neutron polarimetry [5].

Valence states accessed by photo-ejected electrons interact with neighbouring ions when x-rays excite a core resonance. Thus, electronic multipoles in the ground state observed in diffraction are rotationally anisotropic with a symmetry corresponding to the site symmetry of the resonant ion. The distribution in a crystal of spherically symmetric atomic charge defines a space group. In general, strong Bragg spots are absent in diffraction for specific conditions on Miller indices, and absence conditions are listed in an Appendix for the two compounds of interest. Absence conditions can be violated by relatively weak Bragg spots arising from non-spherical atomic charge, which is usually called Templeton-Templeton scattering [36, 37, 38]. Space group forbidden Bragg spots are particularly revealing with regard to fine features of the electronic structure. Tuning the energy of x-rays to an atomic resonance has two obvious benefits. In the first place, there is a welcome enhancement of Bragg spot intensities and, secondly, spots are element specific. There are four scattering amplitudes labelled by photon polarization, two with unrotated and two with rotated states of polarization [13, 38]. Strong Thomson scattering, by spherically symmetric atomic charge, that overwhelms weak signals is absent in rotated channels of polarization. It is allowed in unrotated channels of polarization using a parity-even absorption event, but absent in a parity-odd absorption, e.g., electric dipole - electric dipole (E1-E1) and electric dipole - electric quadrupole (E1-E2) events. Diffraction amplitudes for E1-E1, E2-E2 and E1-E2 events presented here include rotation of the crystal about the reflection vector (azimuthal-angle scan) and they are specific to position multiplicity, Wyckoff letter and symmetry [38]. The range of values of the rank K is fixed by the triangle rule, and K = 0 - 2, K = 1 - 3 and K= 0 - 4 for E1-E1, E1-E2 and E2-E2 events, respectively. Angular rotation symmetry in a crystal can be mirrored in a periodicity of an azimuthal-angle scan, and the valuable property of the diffraction technique depends on the direction of the chosen reflection vector relative to crystal axes.

Magnetic multipoles in x-ray diffraction using E1-E1 or E2-E2 absorption events have an odd rank, i.e., K = 1 or 3 [13, 38]. Whereas, a parity-odd E1-E2 event presents Dirac multipoles with K = 1 - 3. Magnetic contributions to neutron diffraction patterns are often identified in differences between patterns taken at different temperatures, above and below the onset of long-range magnetic order. Greater sensitivity is afforded by exploiting polarization analysis. A spin-flip signal has been used to good effect in measuring weak magnetic signals from long-range order in the pseudo-gap phase of high-temperature superconducting cuprates [39-43].

Bulk magnetic properties of iron tellurate are made exclusively of Dirac multipoles, including monopoles. Although anapoles are forbidden, Dirac multipoles alone are responsible for a whole class of Bragg spots for which we give scattering amplitudes fully informed by symmetry. Axial quadrupoles produce weak (space group forbidden) reflections. They account for a correlation between anapole and orbital degrees of freedom in the magnetic neutron scattering amplitude. An axial secondary order parameter in the spin ladder compound contributes weak Bragg spots. Axial dipoles and anapoles are orthogonal, with anapoles confined to the a-axis in Fig. 2. However, anapoles do not contribute to bulk properties of $SrFe_2S_2O$. A Kerr effect is allowed, unlike iron tellurate.

## II. IRON TELLURATE

The rutile-type structure (space group $P4_2/mnm$) contains chains of edge-sharing octahedra along the crystal c-axis joined through corners in the a-b plane. Trirutile materials ($AB_2X_6$ composition) use a superstructure with a tripling of the rutile unit along the c-axis [44]. Antiferromagnetic chains parallel with the [1, 1, 0] axis, distinguished by the closest metal-metal separation in a trirutile, are a characteristic of the magnetic states [45]. The compound of interest, $Fe_2TeO_6$, is referred to as an "inverted" trirutile, because A and B cations are interchanged with respect to the nominal trirutile [46]. It displays a linear magnetoelectric effect (ME), as does $Cr_2WO_6$ [47, 48]. The mentioned trirutile oxides possess different long-range magnetic structures and magnetic crystal classes, however, with moments parallel (Fe, 4/m'm'm') and perpendicular (Cr, m'mm) to the c-axis [23].

Absence conditions on Miller indices ($h$, $k$, $l$) for Bragg diffraction by $P4_2/mnm$ are listed in an Appendix. The configuration of axial magnetic dipoles depicted in Fig. 1 belongs to the tetragonal centrosymmetric space group $P4_2/m'n'm'$ (No. 136.503 BNS [49]), and ME crystal class 4/m'm'm'. Ferric ions ($Fe^{3+}$, $3d^5$) are in non-centrosymmetric Wyckoff sites 4(e), with site symmetry 2.m'm' comprising two operations $2_z$ (dyad axis of rotation symmetry along the c-axis in Fig. 1) and $m_{xy}'$ (mirror and time-reversal invariances along a diagonal in the ab-plane). Cell lengths a ≈ 4.601 Å and c ≈ 9.087 Å, and $T_N$ ≈ 209 K [23]. There are two independent parameters in the ME coupling (susceptibility) tensor [24]. A piezomagnetic effect is forbidden by anti-inversion in the crystal class.

Properties antiferromagnetic $Fe_2TeO_6$ contrast with those of $MnF_2$ [50], $CoF_2$ and $FeF_2$ that also have the rutile structure $P4_2/mnm$. Magnetic properties of the cited fluorides belong to the centrosymmetric crystal class 4'/mmm'. Absence of anti-inversion in the crystal class means that the ME effect is non-linear, a piezomagnetic effect is allowed, and the classical polarized neutron diffraction technique is available [5, 31, 32].

### A. X-ray diffraction

A universal structure factor of rank K,

$$\Psi^K{}_Q = [\exp(i\boldsymbol{\kappa}\cdot\mathbf{d})\,\langle O^K{}_Q\rangle_{\mathbf{d}}], \qquad (1)$$

determines diffraction amplitudes. An electronic multipole $\langle O^K{}_Q\rangle$ possesses (2K + 1) projections in the interval $-K \leq Q \leq K$, and the complex conjugate obeys $(-1)^Q \langle O^K{}_{-Q}\rangle = \langle O^K{}_Q\rangle^*$. Multipoles for x-ray and neutron scattering abide by the same discrete symmetry requirements but they are different in detail. The structure factor $\Psi^K{}_Q$ is informed of all elements of symmetry in the magnetic space group. In Eq. (1), $\boldsymbol{\kappa}$ is the reflection vector defined by integer Miller indices ($h$, $k$, $l$), and the implied sum is over the four Fe ions in sites $\mathbf{d}$ in a magnetic unit cell. Evaluated for $Fe_2TeO_6$,

$$\Psi^K{}_Q(\text{TET}) = \langle O^K{}_Q\rangle\,[\gamma + \sigma_\pi \sigma_\theta\, \gamma^*]\,[1 + (-1)^n\,(-1)^{h+k+l}], \qquad (2)$$

where the condition Q = 2n on projections flows from the dyad axis of rotation symmetry on the c-axis. Site symmetry demands $\sigma_\pi \sigma_\theta (-1)^{K+n} \langle O^K_{-Q} \rangle = \langle O^K_Q \rangle$. Here, $\sigma_\pi$ and $\sigma_\theta$ are signatures of discrete symmetries of parity and time, with $\sigma_\pi = +1$ (−1) for axial (polar) and $\sigma_\theta = +1$ (−1) for time-even (time-odd, magnetic). The spatial phase $\gamma = \exp(i2\pi zl)$ with $z \approx 0.334$ [23].

Magnetic reflection conditions reported by Kunnmann *et al*. [23] are even $h + k + l$, and odd $h + k + l$ absent. They apply to axial magnetic dipoles. The conditions are reproduced by our structure factor Eq.(2) after setting $\sigma_\pi \sigma_\theta = -1$ and K = 1, and aligning dipoles with the c-axis. Notably, reflections indexed ($h$, $k$, 0) with $\gamma = 1$ are forbidden for all axial multipoles for all $h$, $k$.

Dirac multipoles $\langle G^K_Q \rangle$ possess $\sigma_\pi = -1$ (parity-odd) and $\sigma_\theta = -1$ (magnetic) and they alone are responsible for magnetic Bragg reflections ($h$, $k$, 0). In particular, bulk magnetic properties specified by (0, 0, 0) comprise Dirac multipoles with projections even $n$, including magnetic monopoles. The latter are revealed by the diffraction of light using the electric dipole - magnetic dipole (E1-M1) parity-odd absorption event [8].

Here, we consider Bragg diffraction of hard x-rays enhanced by the Fe K-edge (energy ≈ 7.115 keV and wavelength $\lambda \approx 1.743$ Å) using an electric dipole - electric quadrupole event (E1-E2) [51]. Primary (secondary) photon polarizations parallel and perpendicular to the plane of scattering are labelled $\pi$ ($\pi'$) and $\sigma$ ($\sigma'$), respectively. The four scattering amplitudes for space group forbidden ($h$, 0, 0) with odd $h$ are (apart from numerical factors and radial integrals [38, 52]),

$(\sigma'\sigma) = \cos(\theta) \sin(\psi) \langle G^2_{+2} \rangle''$, $(\pi'\pi) = -\cos(3\theta) \sin(\psi) \langle G^2_{+2} \rangle''$,

$(\pi'\sigma) = -(\sigma'\pi) = \sin(2\theta) \cos(\psi) \langle G^2_{+2} \rangle''$.  ($h$, 0, 0) odd $h$ \hspace{1em} (3)

Rotation of the Fe$_2$TeO$_6$ crystal around the reflection vector is measured by the angle $\psi$, starting from the c-axis normal to the scattering plane. The Bragg angle $\theta$ is determined by $\sin(\theta) = \lambda h/(2a) \approx h\, 0.189$, and $h = 1, 3, 5$ satisfy the Laue condition. To reiterate, Bragg spots ($h$, 0, 0) with odd $h$ are magnetic and solely determined by a Dirac quadrupole $\langle G^2_{+2} \rangle'' = \text{Im.}\, \langle G^2_{+2} \rangle$. No additional information is available from Bragg spots indexed by remaining classes of space group forbidden reflections. Equivalent reflections and absence conditions for space group P4$_2$/mnm are listed in an Appendix for the convenience of the reader.

### B. Neutron diffraction

Unlike multipoles for resonant x-ray diffraction, multipoles in neutron diffraction are strong functions of the magnitude of the reflection vector $\kappa$. Radial integrals $\langle j_c(\kappa) \rangle$ in our axial multipoles $\langle t^K_Q \rangle$ are averages of spherical Bessel functions and integer $c$ is even. Values of $\langle j_0(\kappa) \rangle$ and $\langle j_2(\kappa) \rangle$ for ferric ions are displayed in Fig. 2 of Ref. [53]. By definition, $\langle j_c(0) \rangle = 0$ for $c > 0$, and $\langle j_0(0) \rangle = 1$. Most simple models of magnetic neutron scattering are based on the dipole $\langle t^1 \rangle$ and exclude all but $\langle j_0(\kappa) \rangle$, cf. Eq. (11) [53, 54].

Magnetic and nuclear scattering amplitudes for Fe$_2$TeO$_6$ are 90° out of phase. This finding is expected on the grounds that magnetic symmetry allows the ME effect. Specifically, the identity $\{-(-1)^{K+n} \langle O^K_Q \rangle^*\} = \langle O^K_Q \rangle$ is imposed on axial multipoles by site symmetry, and the multipoles are purely real (imaginary) for odd (even) $K + n$. The Fe axial dipole (K = 1, Q = 0) and quadrupole (K = 2, Q = ±2) are purely real. An equivalent operator [(S×R) R] for the quadrupole shows that it measures the correlation between a spin anapole (S×R) and orbital degrees of freedom [55]. With $n = 1$ in Eq. (2) the quadrupole may contribute to scattering for odd $h + k + l$. Likewise the off-diagonal octupole (K = 3, Q = ±2) that is purely imaginary.

Reflections (0, $k$, $l$) odd $k + l$ are absent in nuclear scattering. The corresponding time-average of the neutron scattering operator $\langle \mathbf{Q} \rangle^{(+)} = (\langle Q_x \rangle^{(+)}, 0, 0)$ [34, 35, 55] is purely imaginary and,

$$\langle Q_x \rangle^{(+)} \approx i e_y e_z \sin(2\pi z l) [\langle t^2_{+2} \rangle' + (1/2) \sqrt{(35/2)} \langle t^3_{+2} \rangle'']. \quad (0, k, l) \text{ odd } k + l \quad (4)$$

is correct at the level of octupoles. A unit vector $\mathbf{e}$ is parallel to the reflection vector and components $e_y \propto k$, $e_z \propto l$. Since $(2\pi z) \approx 120°$ the amplitude Eq. (4) is small for Miller indices $l$ that are multiples of 3. The superscript $^{(+)}$ on the neutron amplitude denotes parity-even. The quadrupole $\langle t^2_{+2} \rangle'$ in Eq. (4) is proportional to the radial integral $\langle j_2(\kappa) \rangle$ that also features in the octupole $\langle t^3_{+2} \rangle''$. It has a broad maximum around $\kappa \approx 6$ Å$^{-1}$ [53].

In general, a polarized neutron diffraction signal $\Delta = \{\mathbf{P} \cdot \langle \mathbf{Q}_\perp \rangle\}$, where $\mathbf{P}$ is polarization of the primary neutrons and $\langle \mathbf{Q}_\perp \rangle = \{\mathbf{e} \times (\langle \mathbf{Q} \rangle \times \mathbf{e})\}$. A spin-flip intensity SF is a measure the magnetic content of a Bragg spot, and SF = $\{|\langle \mathbf{Q}_\perp \rangle|^2 - \Delta^2\}$ when $\mathbf{P} \cdot \mathbf{P} = 1$ and $(\langle \mathbf{Q}_\perp \rangle^* \times \langle \mathbf{Q}_\perp \rangle) = 0$ [41].

Magnetic monopoles are forbidden in $\langle \mathbf{Q}_\perp \rangle^{(-)}$ [55], and the Fe anapole (K = 1) is zero in Fe$_2$TeO$_6$. In consequence, the leading contribution to diffraction by Dirac multipoles is made by the imaginary part of the spin quadrupole, $\langle g^2_{+2} \rangle''$. The associated radial integral is zero in the forward direction of scattering [53]. Consider reflections (0, $k$, $l$) absent in nuclear scattering for odd $k + l$. The component of $\langle \mathbf{Q}_\perp \rangle^{(-)}$ parallel to the a-axis is different from zero, $\langle \mathbf{Q}_\perp \rangle^{(-)} = (\langle Q_{\perp,x} \rangle^{(-)}, 0, 0)$ and, neglecting Dirac multipoles beyond the mentioned quadrupole,

$$\langle Q_{\perp,x} \rangle^{(-)} \approx i e_y \cos(2\pi z l) \langle g^2_{+2} \rangle''. \quad (5)$$

Thus,
$$SF \approx |\langle Q_{\perp,x} \rangle^{(-)}|^2 (1 - P_x^2), \quad (0, k, l) \text{ odd } k + l \quad (6)$$

and the spin-flip signal is zero for $\mathbf{P}$ normal to the reflection vector, and potentially non-zero otherwise. Optimum intensities occur for $l = 0, 3, 6$, etc.

### III. Spin ladder SrFe$_2$S$_2$O

A reduced symmetry of the orthorhombic spin ladder compound compared to tetragonal iron tellurate means fewer constraints on multipoles and diffraction amplitudes, which contain comparatively more multipoles. Diffraction by axial magnetic dipoles superimposes with core Bragg spots, as with iron tellurate, and the magnetic propagation vector = (0, 0, 0) for both antiferromagnets under investigation. Equivalent reflections and absence conditions for the parent structure Pmmn (No. 59) are listed in an Appendix. Cell lengths a ≈ 3.870 Å, b ≈ 9.379 Å and c ≈ 6.307 Å [28].

The magnetic space group Pm'm'n' (No. 59.411, origin choice 2, BNS) of $SrFe_2S_2O$ has only recently been determined [28]. It is centrosymmetric and belongs to the magnetic crystal class m'm'm' that forbids a piezomagnetic effect. The antiferromagnetic configuration of axial dipoles normal to the a-axis below $T_N$ ≈ 216 K is depicted in Fig. 2. The refined magnetic structure implies the magnetic axial dipole along the b-axis to be the primary order parameter, which defines the Pm'm'n' space group. The latter allows the orthogonal axial component along the c-axis as a secondary order parameter coupled to the primary one via a weak interaction, e.g., the Dzyaloshinskii-Moriya interaction. Unlike tetragonal iron tellurate, there are three independent parameters in the diagonal ME coupling tensor. And, the magnetic crystal class m'm'm' is one of three with no bulk axial magnetism that allow the Kerr effect [56]. Anapoles (Dirac dipoles) are another feature of the spin ladder compound not present in iron tellurate.

Ferrous ions $Fe^{2+}$ ($3d^6$) use Wyckoff sites 4(e) with symmetry $m_x$' (mirror and time-reversal invariances along the a-axis in Fig. 2). An electronic structure factor for a reflection vector (h, k, l) with integer Miller indices,

$$\Psi^K_Q(ORH) = \exp(i\pi h/2) \langle O^K_Q \rangle [\beta + \beta^* (-1)^{Q+k}] [\gamma + \sigma_\pi\sigma_\theta \gamma^* (-1)^Q (-1)^{h+k}]. \qquad (7)$$

Site symmetry for ferrous ions demands $\sigma_\pi\sigma_\theta (-1)^K \langle O^K_{-Q} \rangle = \langle O^K_Q \rangle$, while $(-1)^Q \langle O^K_{-Q} \rangle^* = \langle O^K_Q \rangle$ and $\sigma_\pi\sigma_\theta = -1$ (+1) for axial (Dirac) multipoles, respectively. Spatial phases in Eq. (7) are $\beta = \exp(i2\pi yk)$, $\gamma = \exp(i2\pi zl)$ with y ≈ 0.582 and z ≈ 0.378 [28].

Axial multipoles with even (odd) K + Q are purely imaginary (real). The classification is reversed for Dirac multipoles. Cartesian components (x, y, z) of a dipole are derived from $\langle O^1_x \rangle = (\langle O^1_{-1} \rangle - \langle O^1_{+1} \rangle)/\sqrt{2} = -\sqrt{2} \langle O^1_{+1} \rangle'$, $\langle O^1_y \rangle = i(\langle O^1_{-1} \rangle + \langle O^1_{+1} \rangle)/\sqrt{2} = -\sqrt{2} \langle O^1_{+1} \rangle''$, $\langle O^1_z \rangle = \langle O^1_0 \rangle$. Anapoles are normal to the bc-plane that contains axial dipoles. Bulk magnetic properties comprise Dirac multipoles with projections even Q, and include the monopole $\langle G^0 \rangle$ and quadrupoles $\langle G^2 \rangle$.

### A. X-ray diffraction

Diffraction enhanced by a parity-even absorption event is allowed at space group forbidden reflections, in contrast to iron tellurate. The Fe $L_2$ edge has an energy 721 eV, λ/(2b) ≈ 0.917 and k = 1 is allowed in (0, k, 0). The E1-E1 amplitude in the rotated channel of polarization (π'σ) is proportional to the axial moment parallel to the c-axis, and the corresponding multipole is denoted $\langle T^1_z \rangle$ in amplitudes,

$$(\pi'\sigma) = -(\sigma'\pi) = \sin(2\pi y) \cos(\theta) \sin(\psi) \langle T^1_z \rangle, \qquad (0, 1, 0) \qquad (8)$$

with sin(2πy) = − 0.493. Axial octupoles are engaged in diffraction by an E2-E2 event at the Fe K-edge [38, 57, 58] with the result,

$$(\pi'\sigma) = -(\sigma'\pi) = -\sin(2\pi yk)\sin(\psi)[\cos(3\theta)\langle T^1{}_z\rangle + \{3\cos(3\theta)$$
$$+ 5\Phi\sin^2(\psi)\}\langle T^3{}_0\rangle - \sqrt{30}\{\cos(\theta)\cos(2\theta) + \Phi\sin^2(\psi)\}\langle T^3{}_{+2}\rangle'], \,(0, k, 0) \text{ odd } k \quad (9)$$

where $\Phi = \cos(\theta)[2 - 3\cos^2(\theta)]$. E2-E2 amplitudes Eq. (9) are potentially strong for (0, 3, 0), for which $\sin(6\pi y) \approx -1.0$, $\cos(3\theta) \approx 0.662$, $\cos(\theta)\cos(2\theta) \approx 0.811$, and $\Phi \approx -0.737$. Moreover, diffraction by Dirac multipoles using the parity-odd absorption event E1-E2 is negligible for (0, 3, 0), as we shall see. The condition $\sin(\theta) = \lambda k/(2b) \approx k\, 0.093$ is satisfied for $k = 1, 3, 5, 7, 9$. Contributions $\langle T^1{}_z\rangle$ in Eqs. (8) and (9) merit comment, which is found following Eq. (11) in the following subsection.

In discussing diffraction by Dirac multipoles, we continue to explore space group forbidden reflections. The structure factor Eq. (7) is zero for $(h, 0, 0)$ with odd $h$. Projections Q are restricted to odd values for $(0, k, 0)$ with odd $k$, and in the rotated E1-E2 channel,

$$(\pi'\sigma) = -(\sigma'\pi) = \cos(2\pi yk)\sin(2\theta)\sin(\psi)[3\langle G^1{}_{+1}\rangle' + 2\sqrt{5}\langle G^2{}_{+1}\rangle''$$
$$+ \{4 - 15\cos^2(\psi)\}\langle G^3{}_{+1}\rangle' + \sqrt{15}\cos^2(\psi)\langle G^3{}_{+3}\rangle']. \,(0, k, 0) \text{ odd } k \quad (10)$$

As previously mentioned, $\cos(6\pi y) \approx 0.0$. Miller index $k = 7$ is optimum for the pre-factors $\sin(2\theta)$ and $\cos(2\pi yk)$. Note that a quadrupole is absent from the E2-E2 amplitude, because magnetic parity-even diffraction uses odd rank multipoles alone.

### B. Neutron diffraction

For $(0, k, 0)$ with odd $k$, $\langle \mathbf{Q}\rangle^{(+)} = (0, 0, \langle Q_z\rangle^{(+)})$ is purely imaginary, as expected, and,

$$\langle Q_z\rangle^{(+)} \approx i\sin(2\pi yk)[\langle t^1{}_z\rangle - (2/\sqrt{3})\langle t^2{}_{+2}\rangle''$$
$$- (1/4)\sqrt{7}\{\langle t^3{}_0\rangle + 2\sqrt{(5/6)}\langle t^3{}_{+2}\rangle'\}]. \,(0, k, 0) \text{ odd } k \quad (11)$$

The result is correct for all multipoles up to and including K = 3. The presence in Eq. (11) of the axial magnetic dipole appears, at first sight, to be at odds with the conclusion by Guo *et al.* [28] that chemical and magnetic dipole structures match. Two elementary resolutions are; (i) the component of the axial dipole $\langle t^1{}_z\rangle$ parallel to the c-axis is zero or (ii) $\langle t^1{}_z\rangle$ is too small to be detected with the resolution available to Guo *et al.* [28]. As it stands, dipoles $\langle t^1{}_y\rangle$ constitute the primary order parameter, with $\langle t^1{}_z\rangle$ contributing a comparatively weak secondary order parameter. Note, while Fe site symmetry $m_x'$ allows $\langle t^1{}_z\rangle$ it can be identically zero. Returning to $Fe_2TeO_6$, the predicted magnetic amplitude Eq. (4) for non-core Bragg spots has the expected structure, i.e., no axial dipoles. Intensity of a Bragg spot = $|\langle Q_z\rangle^{(+)}|^2$ since $\langle \mathbf{Q}\rangle^{(+)}$ and **e** are orthogonal.

The result in Eq. (11) holds for $(h, 0, 0)$ with $h = (2n + 1)$ on making one change. The spatial phase factor $\sin(2\pi yk)$ is replaced by $(-1)^n$.

An approximation for the axial dipole in Eq. (11),

$$\langle \mathbf{t}^1\rangle \approx (\langle \mathbf{\mu}\rangle/3)[\langle j_0(\kappa)\rangle + \langle j_2(\kappa)\rangle(g-2)/g], \quad (12)$$

is often used [34, 55]. Here, the magnetic moment ⟨**μ**⟩ = g ⟨**S**⟩ and the orbital moment ⟨**L**⟩ = (g − 2) ⟨**S**⟩.

Regarding diffraction by Dirac multipoles, ⟨**Q**⊥⟩$^{(-)}$ = (0, 0, ⟨$Q_{\perp,z}$⟩$^{(-)}$) for (0, k, 0) with odd k. At an accuracy that includes anapoles and quadrupoles,

$$\langle Q_{\perp,z}\rangle^{(-)} \approx - i \cos(2\pi y k) \langle g^2_{+1}\rangle''. \quad (0, k, 0) \text{ odd } k \quad (13)$$

The spatial phase factor is zero for k = 3, to a good approximation, while Eq.(11) for the axial contribution to diffraction is optimal.

### IV. DISCUSSION

In summary, we have identified Dirac multipoles in two antiferromagnetic iron based magnetoelectric (ME) compounds cited in Table I, and depicted in Figs. 1 and 2. They are polar and magnetic (time-odd), and represent the scalar, dipole and quadrupole in a decomposition of the third-order tensor that couples electric and magnetic fields in a Landau free-energy. Findings are informed by symmetry in previously established magnetic space groups for iron tellurate [23] and a spin ladder compound [28]. A linear ME effect in iron tellurate has been thoroughly investigated, starting in 1972 with a study reported by Buksphan *et al*. [24]. Not so for the spin ladder $SrFe_2S_2O$, to the best of our knowledge. However, its magnetic space group belongs to the crystal class m'm'm' that allows a linear ME effect. Even more interestingly, the magnetic crystal class is one of three with no bulk axial magnetism that allow the Kerr effect [56]. Dirac multipoles form bulk magnetization in both iron based compounds, and it includes monopoles (often called magnetic charges).

The published magnetic space group Pm'm'n' for the ladder compound implies magnetic axial dipoles along the b-axis in Fig 2 form a primary order parameter. This space group also allows an orthogonal axial component along the c-axis. The latter seems to be a secondary order parameter weakly coupled to the primary one and therefore very small to be reliably detected in a diffraction experiment on a powder sample [28]. Usually, orthogonal magnetic components due to spin canting imposed by the Dzyaloshinskii-Moriya interaction are roughly about 0.01-0.02 $\mu_B$, which is almost an order of magnitude smaller than the sensitivity of good quality powder diffraction data [59].

The defining property of magnetoelectric space groups is the presence of anti-inversion ($\bar{1}'$) among invariance requirements. It has global influence on diffraction amplitudes for antiferromagnetic materials, which is illustrated by our results. Nuclear and magnetic contributions to the neutron scattering amplitude are forced to differ by 90º. In the diffraction of x-rays with energy matching an iron resonance charge and magnetic contributions have a like phase. In consequence, Bragg diffraction patterns are independent of helicity carried by primary x-rays.

Special attention in our calculations is given to weak Bragg spots stemming from angular anisotropy in magnetic distributions. Such reflections violate absence conditions derived from diffraction patterns generated by nuclei or spherical distributions of electronic charge, which are listed in an Appendix together with reflection equivalences. Weak Bragg spots reveal fine features of the distribution of magnetization in a crystal. They feature axial quadrupoles, the correlation between a spin anapole and orbital degrees of freedom, anapoles and Dirac quadrupoles. Calculated x-ray diffraction patterns include the effect of rotating the crystal about the reflection vector (an azimuthal angle scan) that reveals magnetic symmetries.

**ACKNOWLEDGEMENTS** Dr K. S Knight and Dr D. D. Khalyavin were generous with guidance on crystal and magnetic symmetries in calculations of Bragg diffraction patterns. Professor S. P. Collins commented on the feasibility of proposed x-ray Bragg diffraction experiments.

---

TABLE I. Properties of antiferromagnetic iron based ME compounds. The propagation vector = (0, 0, 0), space groups are centrosymmetric, the piezomagnetic effect is forbidden, and $\alpha_{ij}$ are diagonal components of the third-order ME coupling tensor. Magnetic ions occupy sites 4(e). Saturation magnetic moments are in units of the Bohr magneton $\mu_B$. Our investigation of the orthorhombic compound applies also to $SrFe_2Se_2O$ with a different Néel temperature $T_N$ and cell edges [28].

| $Fe_2TeO_6$ [23, 24] | $SrFe_2S_2O$ [28] |
|---|---|
| tetragonal | orthorhombic |
| $P4_2/m'n'm'$ (No. 136.503) | $Pm'm'n'$ (No. 59.411) |
| $4/m'm'm'$ | $m'm'm'$ |
| $Fe^{3+}(3d^5)$ | $Fe^{2+}(3d^6)$ |
| $\alpha_{11}, \alpha_{11}, \alpha_{33}$ | $\alpha_{11}, \alpha_{22}, \alpha_{33}$ |
| $T_N \approx 209$ K | $T_N \approx 216$ K |
| $\mu \approx 4.2\ \mu_B$ | $\mu \approx 3.3\ \mu_B$ |
| monopole | monopole |
| no anapole | anapole |
| no Kerr effect | Kerr effect [56] |

---

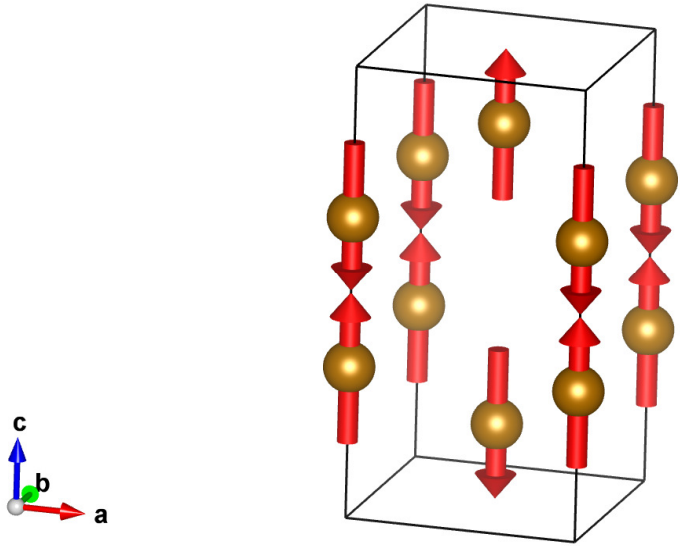

FIG. 1. Configuration of ferric axial magnetic dipoles in the inverted trirutile $Fe_2TeO_6$ [23]. Reproduced from MAGNDATA [29].

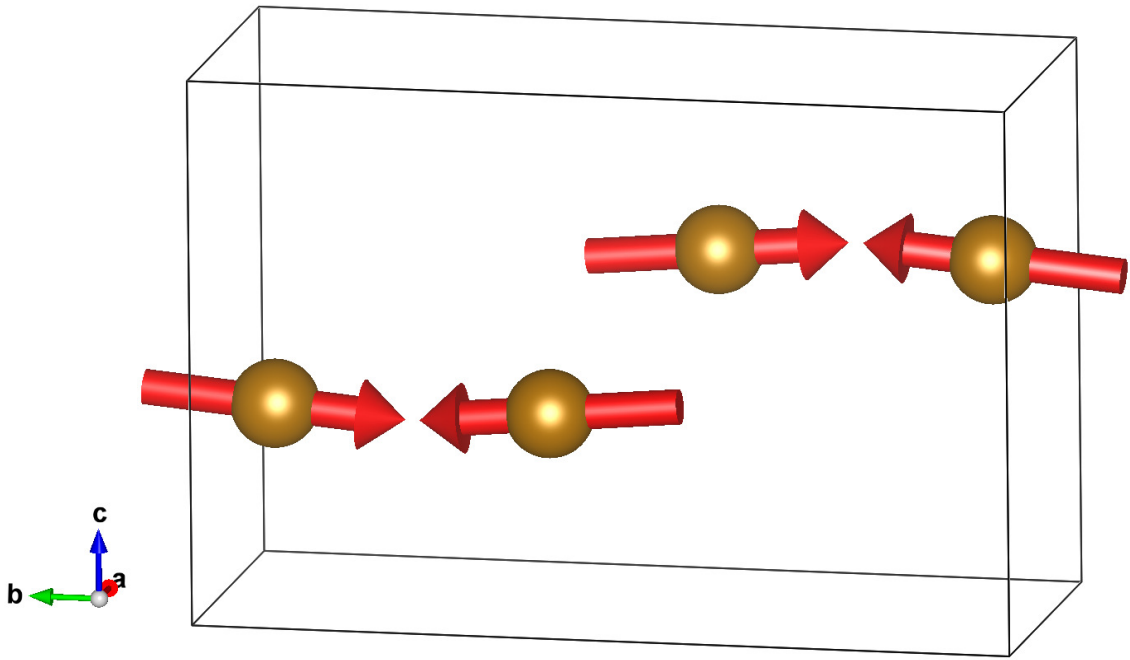

FIG. 2. Configuration of ferrous axial magnetic dipoles in the orthorhombic spin ladder compound $SrFe_2S_2O$ [28]. Reproduced from MAGNDATA [29].

## Appendix

Equivalent reflections for P4$_2$/mnm (No. 136)

(1) $h, k, l$ (2) $-h, -k, l$ (3) $-k, h, l$ (4) $k, -h, l$

(5) $-h, k, -l$ (6) $h, -k, -l$ (7) $k, h, -l$ (8) $-k, -h, -l$

(9) $-h, -k, -l$ (10) $h, k, -l$ (11) $k, -h, -l$ (12) $-k, h, -l$

(13) $h, -k, l$ (14) $-h, k, l$ (15) $-k, -h, l$ (16) $k, h, l$

$(0, 0, l)$ odd $l$ absent; $(h, 0, 0)$ odd $h$ absent; $(0, k, l)$ odd $k + l$ absent

Equivalent reflections for Pmmn (No. 59)

(1) $h, k, l$ (2) $-h, -k, l$ (3) $-h, k, -l$ (4) $h, -k, -l$

(5) $-h, -k, -l$ (6) $h, k, -l$ (7) $h, -k, l$ (8) $-h, k, l$

$(h, k, 0)$ odd $h + k$ absent; $(h, 0, 0)$ odd $h$ absent; $(0, k, 0)$ odd $k$ absent